\documentclass[pra,twocolumn,a4paper,showpacs,floatfix]{revtex4} 

\usepackage{graphicx}
\usepackage{amsmath,amssymb}
\newcommand{\figwidth}{0.95\columnwidth}

\begin{document}

\title{Electronic Transport in DNA}

\author{Daphne Klotsa, Rudolf A.\ R\"{o}mer, Matthew S.\ Turner}
\affiliation{Physics Department and Centre for Scientific
Computing, University of Warwick, Coventry CV4 7AL, U.K.}

\date{$Revision: 1.45 $, compiled \today}

\begin{abstract}
  We study the electronic properties of DNA by way of a tight-binding
  model applied to four particular DNA sequences. The charge transfer
  properties are presented in terms of localisation lengths, crudely speaking the length over which electrons travel. Various types of disorder, including random potentials, are employed to account for different real
  environments. We have performed calculations on poly(dG)-poly(dC),
  telomeric-DNA, random-ATGC DNA and $\lambda$-DNA. We find that random
  and $\lambda$-DNA have localisation lengths allowing for electron
  motion among a few dozen base pairs only.  A novel enhancement of
  localisation lengths is observed at particular energies for an
  increasing binary backbone disorder. We comment on the possible
  biological relevance of sequence dependent charge transfer in DNA.
\end{abstract}


\maketitle

\section{Introduction}
\label{sec-introduction}

The question of whether DNA conducts electric charges is intriguing to
physicists and biologists alike. The suggestion that electron transfer/transport in DNA might be
biologically important has triggered a series of experimental and
theoretical investigations
\cite{MurAJG93,BerBR00,WanFSB00,DelB03,OneDB04,EndCS04}. Processes that
possibly use electron transfer include the function of DNA damage
response enzymes, transcription factors or polymerase co-factors all of
which play important roles in the cell \cite{AlbBLR94}. Indeed there is
direct evidence \cite{BooLCD03} that MutY --- a DNA base excision repair
enzyme with an [4Fe4S]$^+$ cluster of undetermined function ---
takes part in some kind of electron transfer as part of the DNA repair
process \cite{OneF93,RetHBL93}. This seems consistent with studies in
which an electric current is passed through DNA revealing that damaged
regions have significantly different electronic behaviour than healthy
ones \cite{BooLCD03}.

For physicists, the continuing progress
of nanotechnologies and the consequent need for further size
miniaturisation makes the DNA molecule an excellent candidate for
molecular electronics \cite{CunCPD02,GarABG01,RakAPK01,BhaBB03}. DNA
might serve as a wire, transistor, switch or rectifier depending on its
electronic properties \cite{DekR01,PorCD04,EndCS04}.

In its natural environment, DNA is always in liquid solution and
therefore experimentally one can study the molecule either in solution
or in artificially imposed dry environments. In solution experiments DNA
can be chemically processed to host a donor and an acceptor molecule at
different sites along its long axis. Photo-induced charge transfer rates
can then be measured whilst the donor/acceptor molecules, the distance
and the sequence of DNA that lies between them are varied.  The
reactions are observed to depend on the type of DNA used, the
intercalation, the integrity of the intervening base pair stack and,
albeit weakly, on the molecular distance
\cite{BerBR00,OneDB04,BooLCD03,DelB03,TreHB01}.

Direct conductivity measurements on dry DNA have also been preformed in
the past few years. The remarkable diversity that characterises the
results seems to arise from the fact that many factors need to be
experimentally controlled. These include methods for DNA alignment and
drying, the nature of the devices used to measure the conductivity, the
type of metallic contacts and the sequence and length of the DNA.  DNA
has been reported to be an insulator \cite{BraESB98}, an ohmic conductor
\cite{RakAPK01,FinS99,NakGSO03,AsaT00,OkaKTS98} and a semiconductor
\cite{PorBVD00}. Theoretically, single-step super exchange
\cite{MurAJG93} and multi-step hopping \cite{BixGWL99} models have
provided interpretations of solution experiments. For experiments in dry
DNA, several additional approaches such as variable range hopping
\cite{YuS01}, one-dimensional quantum mechanical tight-binding models
\cite{CunCPD02,Roc03,ZhaU04a,ZhaU04b,RocBMK03,WanLS04} and non-linear
methods \cite{CueS04,Pey04} have also been proposed.

Despite the lack of a consistent picture for the electronic properties
of DNA, one conclusion has been established: the environment of the DNA
impacts upon its structural, chemical and thus probably also electronic
properties.  Both theoretical and experimental studies show that the
temperature and the type of solution surrounding DNA have a significant
effect on its structure and shape \cite{YuS01,BarCJL01,BruGOR00}.  The
effect of the environment is a key one to this report, where the
environmental fluctuations are explicitly modelled as providing
different types of disorder.

In this work, we focus on whether DNA, when treated as a quantum wire in
the fully coherent low-temperature regime, is conducting or not. To this
end, we study and generalise a tight-binding model of DNA which has been
shown to reproduce experimental \cite{CunCPD02} as well as {\em
  ab-initio} results \cite{DavI04}. A main feature of the model is the
presence of sites which represent the sugar-phosphate backbone of DNA
but along which no electron transport is permissible. We measure the
``strength'' of the electronic transport by the {\em localisation
  length} $\xi$, which roughly speaking parametrises whether an electron
is confined to a certain region $\xi$ of the DNA (insulating behaviour)
or can proceed across the full length $L$ ($\leq \xi$) of the DNA
molecule (metallic behaviour).

Sections \ref{sec-models}--\ref{sec-localization} introduce our models
and the numerical approach.  In section \ref{sec-results-clean}, we
show that DNA sequences with different arrangement of nucleotide bases
Adenine (A), Cytosine (C), Guanine (G) and Thymine (T) exhibit different
$\xi$'s when measured, e.g.\  as function of the Fermi energy $E$.  The
influence of external disorder, modelling variants in the solution,
bending of the DNA molecule, finite-temperature effects, etc., is
studied in section \ref{sec-results-disordered} where we show that,
surprisingly, the models support an increase of $\xi$ when disorder is
increased. We explain that this effect is linked to the existence of the
backbone sites.

\section{Tight-binding models for DNA with a gap in the spectrum}
\label{sec-models}

\subsection{The Fishbone model}
\label{sec-fishbone}

\begin{figure}
  \centering
  \includegraphics[width=\figwidth]{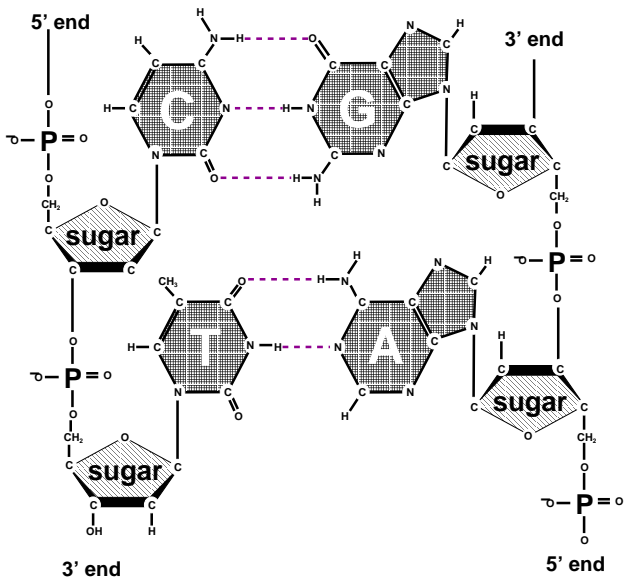}
  \caption{\label{fig-DNA}
    The chemical composition of DNA with the four bases Adenine,
    Thymine, Cytosine, Guanine and the backbone. The backbone is made of
    phosphorylated sugars shown in yellow and brown.}
\end{figure}

DNA is a macro-molecule consisting of repeated stacks of bases formed by
either AT (TA) or GC (CG) pairs coupled via hydrogen bonds and held in
the double-helix structure by a sugar-phosphate backbone. In Fig.\ 
\ref{fig-DNA}, we show a schematic drawing.
%
In most models of electronic transport \cite{CunCPD02,Zho03} it has been
assumed that the transmission channels are along the long axis of the
DNA molecule \cite{perpendiculartolongaxis} and that the conduction path
is due to $\pi$-orbital overlap between consecutive bases
\cite{TreHB01}; density-functional calculations \cite{PabMCH00} have
shown that the bases, especially Guanine, are rich in $\pi$-orbitals.
Quantum mechanical approaches to the problem mostly use strictly
one-dimensional (1D) tight-binding models
\cite{Roc03,ZhaU04a,ZhaU04b,RocBMK03,WanLS04}.

\begin{figure}
  \centering
  \includegraphics[width=\figwidth]{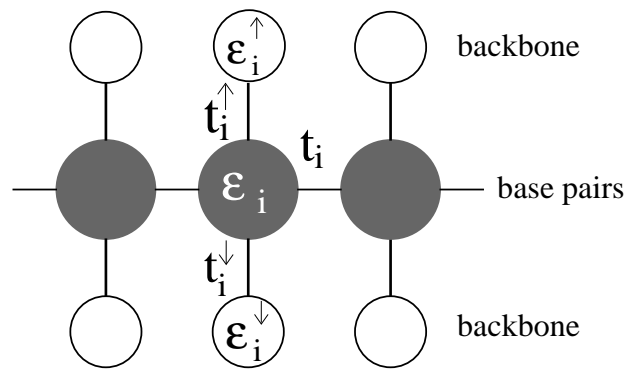}
  \caption{\label{fig-fishbone}
    The fishbone model for electronic transport along DNA corresponding
    to the Hamiltonian given in Eq.\ (\ref{eq-ham1D}). Lines denote
    hopping amplitudes and circles give the central (grey) and backbone
    (open) sites.}
\end{figure}

Of particular interest to us is a quasi-1D model \cite{CunCPD02} which
includes the backbone structure of DNA explicitly and exhibits a
semiconducting gap.
This {\em fishbone model}, shown in Fig.\ \ref{fig-fishbone}, has one
central conduction channel in which individual sites represent a
base-pair; these are interconnected and further linked to upper and
lower sites, representing the backbone, but are \emph{not}
interconnected along the backbone. Every link between sites implies the
presence of a hopping amplitude. The Hamiltonian for the fishbone model
$(H_F)$ is given by:
\begin{eqnarray}
H_F &=& \sum_{i=1}^{L}
 \sum_{q=\uparrow,\downarrow} \left(
    -t_{i} |i \rangle \langle i+1|-t_i^q |i,q \rangle \langle i|
    \right.
 \nonumber \\
 & &
\left.+ \varepsilon_i |i \rangle \langle i| + \varepsilon_i^q |i,q
\rangle \langle i,q| \right) + h.c. \label{eq-ham1D}\label{eq-fishbone}
\end{eqnarray}
where $t_{i}$ is the hopping between nearest-neighbour sites $i,i+1$
along the central branch, $t_i^q$ with $q=\uparrow, \downarrow$ gives
the hopping from each site on the central branch to the upper and lower
backbone respectively. Additionally, we denote the onsite energy at each
site along the central branch by $\varepsilon_i$ and the onsite energy
at the sites of the upper and lower backbone is given by
$\varepsilon_i^q$, with $q=\uparrow\downarrow$. $L$ is the number of
sites/bases in the sequence.
%
The model (\ref{eq-fishbone}) clearly represents a dramatic
simplification of DNA. Nevertheless, in Ref.\ \cite{CunCPD02} it had
been shown that this model when applied to an artificial sequence of
repeated GC base pairs, poly(dG)-poly(dC) DNA, reproduces experimental
data current-voltage measurements when $t_{i}=0.37 e$V and
$t_i^q=0.74 e$V are being used. Therefore, we will assume $t_i^q = 2
t_{i}$ and set the energy scale by $t_{i}\equiv 1$ for hopping
between GC pairs. In what follows we will adopt energy
units in which $eV=1$ throughout.

For natural DNA sequences, we need to know how the hopping amplitudes
vary as the electron moves between like pairs, i.e.\ from GC to GC or
from AT to AT, and unlike pairs, i.e., from GC to AT and vice versa.  We
choose $t_{i}=1$ between identical and matching bases (e.g.\ AT/TA,
GC/CG). Assuming that the wavefunction overlap between consecutive bases
along the DNA strand is weaker between unlike and non-matching bases
(AT/GC, TA/GC, etc.) we thus choose $1/2$.

\subsection{The Ladder model}
\label{sec-ladder}

\begin{figure}
  \centering
   \includegraphics[width=\figwidth]{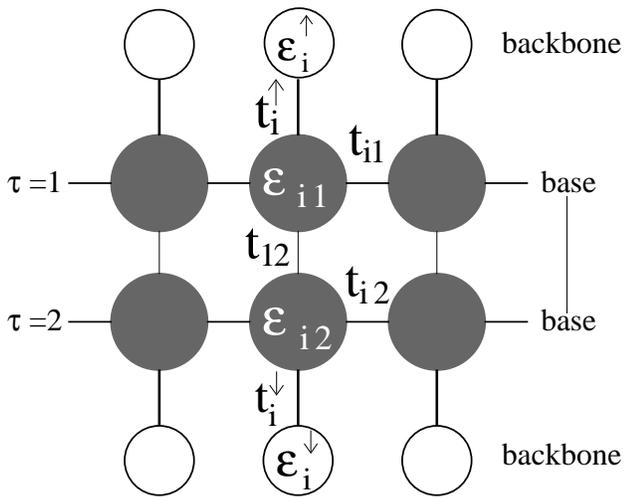}
  \caption{\label{fig-ladder}
    The ladder model for electronic transport along DNA. The model
    corresponds to the Hamiltonian (\ref{eq-ladder}).}
\end{figure}

We performed semi-empirical calculations on DNA base pairs and stacks
using the SPARTAN quantum chemistry software package \cite{Spartan}. The
results have shown that the relevant electronic states of DNA
(highest-occupied and lowest-unoccupied molecular orbitals with and
without an additional electron) are localised on one of the bases of a
pair only. The reduction of the DNA base-pair architecture into a single
site per pair, as in the fishbone model (\ref{eq-fishbone}), is
obviously a highly simplified approach.  As an improvement on this we
model each base as a distinct site where the base pair is then weakly
coupled by the hydrogen bonds. The resulting 2-channel model is shown in
Fig.\ \ref{fig-ladder}. This {\em ladder} model is a planar projection
of the structure of the DNA with its double-helix unwound. We note that
results for electron transfer also suggest that the transfer proceeds
preferentially down one strand \cite{KelB99}.
There are two central branches, linked with one another, with
interconnected sites where each represents a complete base and which are
additionally linked to the upper and lower backbone sites. The backbone
sites as in the fishbone model are not interconnected. The Hamiltonian
for the ladder model is given by
\begin{eqnarray}
H_{L} &=& \sum_{i=1}^{L} \left[
 \sum_{\tau=1,2}
    \left( t_{i,\tau}|i,\tau\rangle \langle i+1,\tau| + 
    \varepsilon_{i,\tau} |i,\tau\rangle \langle i,\tau| \right) \right.
 \nonumber \\
 & &  \mbox{} + \sum_{q=\uparrow,\downarrow}
    \left( t_i^q |i,\tau\rangle \langle i,q(\tau)|+
    \varepsilon_i^q|i,q\rangle \langle i,q| \right) 
\nonumber \\
 & & \mbox{ }
+ t_{1,2}|i,1\rangle \langle i,2| 
\Big]
 + h.c. \label{eq-ham2D}\label{eq-ladder}
\end{eqnarray}
where $t_{i,\tau}$ is the hopping amplitude between sites along each
branch $\tau=1$, $2$ and $\varepsilon_{i,\tau}$ is the corresponding
onsite potential energy. $t_i^q$ and and $\varepsilon_i^q$ as before
give hopping amplitudes and onsite energies at the backbone sites. Also,
$q(\tau)=\uparrow, \downarrow$ for $\tau=1, 2$, respectively. The new
parameter $t_{12}$ represents the hopping between the two central
branches, i.e., perpendicular to the direction of conduction. SPARTAN
results suggest that this value, dominated by the wave function overlap
across the hydrogen bonds, is weak and so we choose $t_{12}= 1/10$.

\subsection{Including disorder}
\label{sec-disorder}

In order to study the transport properties of DNA, we could now either
use artificial DNA (poly(dG)-poly(dC) \cite{PorBVD00}, random sequences
of A,T,G,C \cite{PenBGH92,YamSHA04}, etc.) or natural DNA (bacteriophage
$\lambda$-DNA \cite{PabMCH00}, etc.). The biological content of the
sequence would then simply be encoded in a specific sequence of hopping
amplitudes $1$ and $1/2$ between like and unlike base-pair sequences.
However, in vivo and most experimental situations, DNA is exposed to
diverse environments and its properties, particularly those related to
its conformation, can change drastically depending on the specific
choice.  The solution, thermal effects, presence of binding and
packaging proteins and the available space are factors that alter the
structure and therefore the properties that one is measuring
\cite{YuS01,BarCJL01}.  Clearly, such dramatic changes should also be
reflected in the electronic transport characteristics. Since it is
precisely the backbone that will be most susceptible to such influences,
we model such environmental fluctuations by including variations in the
onsite potentials $\varepsilon_{i,q}$.

Different experimental situations will result in a different
modification of the backbone electronic structure, and we model this by
choosing different distribution functions for the onsite potentials,
ranging from uniform disorder $\varepsilon_{i,q}\in [-W/2,W/2]$, to
Gaussian disorder and on to binary disorder $\varepsilon_{i,q}=
\pm W/2$. $W$ is a measure for the strength of the disorder in all
cases. Particularly the binary disorder model can be justified by the
localisation of ions or other solutes at random positions
along the DNA strand \cite{BarCJL01}.

\subsection{Effective models and the energy gap}
\label{sec-effective}

Due to the non-connectedness of the backbone sites along the DNA
strands, the models (\ref{eq-fishbone}) and (\ref{eq-ladder}) can be
further simplified to yield models in which the backbone sites are
incorporated into the electronic structure of the DNA. The effective
fishbone model is then given by
\begin{eqnarray}
\tilde{H}_F &=& \sum_{i=1}^{L}
    -t_{i} |i \rangle \langle i+1| + h.c. 
 \nonumber \\
& & \mbox{ } + \left[\varepsilon_i - 
\sum_{q=\uparrow,\downarrow}
\frac{\left(t_{i}^{q}\right)^{2}}{\varepsilon_{i}^{q} - E}\right] 
|i \rangle \langle i| \quad .
\label{eq-fishbone-effective}
\end{eqnarray}
Similarly, the effective ladder model reads as
\begin{eqnarray}
\tilde{H}_{L} &=& \sum_{i=1}^{L}
  t_{1,2}|i,1\rangle \langle i,2| +
\sum_{\tau=1,2}
    t_{i,\tau}|i,\tau\rangle \langle i+1,\tau| 
 \nonumber \\
& &
+ 
\left[ \varepsilon_{i,\tau} - 
 \frac{\left(t_{i}^{q(\tau)}\right)^{2}}{\varepsilon_{i}^{q(\tau)} - E}
\right]
|i,\tau\rangle \langle i,\tau|
 \nonumber \\
& & \mbox{ }+ h.c. \quad . \label{eq-ladder-effective}
\end{eqnarray}
In these two models, the backbone has been incorporated into an
energy-dependent onsite potential on the main DNA sites. This
re-emphasises that the presence of the backbone influences the local
electronic structure on the DNA bases and similarly, any variation in
the backbone disorder potentials $\varepsilon_{i}^{\uparrow,\downarrow}$
will results in a variation of {\em effective} onsite potentials as
given in the brackets of Eqs.\ (\ref{eq-fishbone-effective}) and
(\ref{eq-ladder-effective}).

Both models allow to quickly calculate the gap of the completely ordered
system (all onsite potentials zero) by assuming that the lowest-energy
state $\psi=\sum_{i} \psi_{i(,\tau)} |i(,\tau)\rangle$ in each band
corresponds to constant $\psi_i$ ($\psi_{i,\tau}$) whereas for the
highest-energy states, a checker-board pattern is obtained with
$\psi_{i}=\psi_{i+1}$ ($\psi_{i,\tau}=-\psi_{i+1,\tau}$,
$\psi_{i,1}=-\psi_{i,2}$). For the fishbone model, this shows that,
e.g.\ $E_{{\rm min},\mp}=-
t_{i}\mp\sqrt{t_{i}^2+t_{i,\uparrow}^2+t_{i,\downarrow}^{2}}$ and
$E_{{\rm max},\mp}=
t_{i}\mp\sqrt{t_{i}^2+t_{i,\uparrow}^2+t_{i,\downarrow}^{2}}$.
For the chosen set of hopping parameters for
(\ref{eq-fishbone-effective}) and (\ref{eq-ladder-effective}), this
gives $E_{{\rm min},\mp}= -4, 2$ and $E_{{\rm max},\mp}= -2, 4$ for the
fishbone model and $E_{{\rm min},\mp}\approx -3.31, 1.21$ and $E_{{\rm
    max},\mp}= -1.21, 3.31$ for the ladder model.

\section{The numerical approach and localisation}
\label{sec-localization}

There are several approaches suitable for studying the transport
properties of the models (\ref{eq-fishbone}) and (\ref{eq-ladder}) and
these can be found in the literature on transport in solid state
devices, or, perhaps more appropriately, quantum wires. Since the
variation in the sequence of base pairs precludes a general solution, we
will use two methods well-known from the theory of disordered systems
\cite{RomS03}.

The first method is the iterative transfer-matrix method (TMM)
\cite{PicS81a,PicS81b,MacK83,KraM93,Mac94} which allows us in principle
to determine the localisation length $\xi$ of electronic states in
systems with cross sections $M=1$ (fishbone) and $2$ (ladder) and length
$L \gg M$, where typically a few million sites are needed for $L$ to
achieve reasonable accuracy for $\xi$. However, in the present
situation we are interested in finding $\xi$ also for viral DNA strands
of typically only a few ten thousand base-pair long sequences.  Thus in
order to restore the required precision, we have modified the conventional
TMM and now perform the TMM on a system of fixed length $L_0$. This
modification has been previously used \cite{FraMPW95,RomS97b,NdaRS04}
and may be summarised as follows:
After the usual forward calculation with a global transfer matrix ${\cal
  T}_{L_0}$, we add a backward calculation with transfer matrix ${\cal
  T}^{\rm b}_{L_0}$. This forward-backward-multiplication procedure is
repeated $K$ times. The effective total number of TMM multiplications is
$L_{\rm }=2KL_0$ and the global transfer-matrix is ${\tau}_{L_{\rm }} =
\left( {\cal T}^{\rm b}_{L_0} {\cal T}_{L_0}\right)^K$. It can be
diagonalised as for the standard TMM with $K\rightarrow \infty$ to give
${\tau}^{\dagger}_{L_{\rm }} {{\tau}_{L_{\rm }}} \rightarrow \exp[ {\rm
  diag}(4KL_0/\xi_{\tau})]$ with $\tau=1$ or $\tau= 1, 2$ for fishbone
and ladder model, respectively. The largest $\xi_{\tau} \forall \tau$
then corresponds to the localisation lengths of the electron on the DNA
strand and will be measured in units of the DNA base-pair spacing
($0.34$ nm).

The second method that we will use is the recursive Green function
approach pioneered by MacKinnon \cite{Mac80,Mac85}. It can be used to
calculate the dc and ac conductivity tensors and the density of states
(DOS) of a $d$-dimensional disordered system and has been adopted to
calculate all kinetic linear-transport coefficients such as
thermoelectric power, thermal conductivity, Peltier coefficient and
Lorentz number \cite{RomMV03}.

The main advantage of both methods is that they work reliably (i) for
short DNA strands ranging from 13 (DFT studies \cite{PabMCH00}) base
pairs up to 30 base pairs length which are being used in the nanoscopic
transport measurements \cite{DavI04} as well as (ii) for somewhat longer
DNA sequences as modelled in the electron transfer results and (iii)
even for complete DNA sequences which contain, e.g.\ for human
chromosomes up to 245 million base pairs \cite{AlbBLR94}.

\section{DNA sequences}
\label{sec-dna}

The exact arrangement of the four bases A, T, G, C determines the nature
and function of its associated DNA strand such as the chemical
composition of the proteins which are encoded. While previous studies
have aimed to elucidate whether DNA conducts at all, we shall also focus
our attention to investigate how different DNA sequences, be they
artificial or naturally occurring, ``conduct'' charge differently. Thus
we study a set of different DNA.

A convenient starting point for most electronic transport studies
\cite{PorCD04} is the aforementioned poly(dG)-poly(dC) sequence, which
corresponds to a simple repetition of a GC (or CG) pair. Note that
within our models, there is no difference between GC and CG pairs.
Although not occurring naturally, such sequences can be synthesised
easily. Another convenient choice of artificial DNA strand is a simple
{\em random} sequence of the four bases, which we construct with equal
probability for all 4 bases.  However, they are not normally used in
experiments.

As DNA samples existing in living organisms, we shall use $\lambda$-DNA
of the bacteriophage virus \cite{lambda} which has a sequence of 48502
base pairs. It corresponds to a bacterial virus and is biologically very
well characterised. We also investigate the $29728$ bases of the SARS
virus \cite{sars}.
Telomeric DNA is a particular buffer part at the beginning and ends of
of DNA strands for eukaryote cells \cite{AlbBLR94}. In mammals it is a
Guanine rich sequence in which the pattern TTAGGG is repeated over
thousands of bases. Its length is known to vary widely between species
and individuals but we assume a length of 6000 base-pairs.
Last, we show some studies of centromeric DNA for chromosome 2 of yeast
with 813138 base pairs \cite{cen2}. This DNA is also reportedly rich in G
bases and has a high rate of repetitions which should be favourable for
electronic transport.

Initially, we will compute transport properties for complete DNA
sequences, i.e.\ including and not differentiating between coding and
non-coding sequences (this distinction applies to the naturally
occurring DNA strands only). However, we will later also study the
difference between those two different parts of a given DNA. We
emphasise that while non-coding DNA suffers from the label of ``junk'',
it is now known to play several important roles in the functioning of
DNA \cite{AlbBLR94}.

Before leaving the description of our DNA sequences, we note that
occasionally, we show results for ``scrambled'' DNA. This is DNA with
the same number of A, T, C, G bases, but with their order randomised.
Clearly, such sequences contain the same set of electronic potentials
and hopping variations, but would perform quite differently if released
into the wild. A comparison of their transport properties with those
from the original sequence thus allows to measure how important the
exact fidelity of a sequence is.

\section{Results for clean DNA}
\label{sec-results-clean}

\begin{figure}
  \centering
  \includegraphics[width=\figwidth]{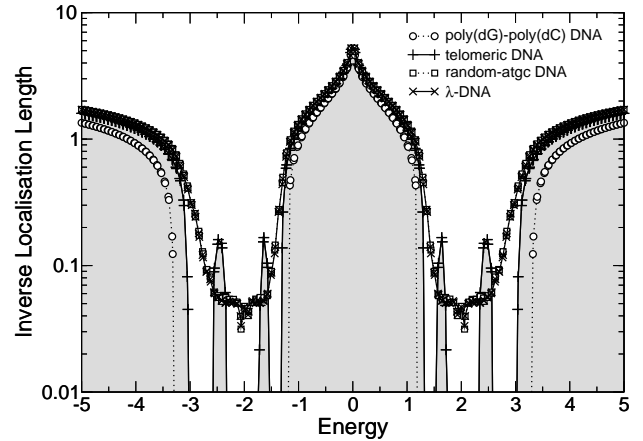}
  \caption{\label{fig-inverse2d}
    Plot of the inverse localisation lengths $\xi$ as a function of
    Fermi energy for the ladder model (\ref{eq-ladder-effective} and
    four DNA sequences as well as for the fishbone model with a
    poly(dG)-poly(dC) sequence. The data for telomeric DNA has been
    shaded for clarity. Lines are guides to the eye only.}
\end{figure}

\begin{figure}
  \centering
  \includegraphics[width=\figwidth]{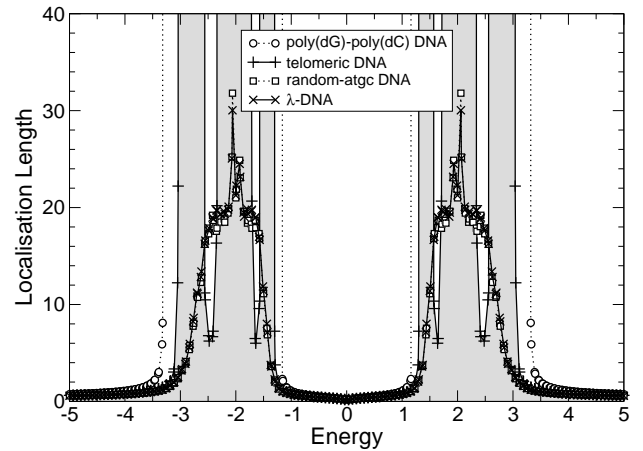}
  \caption{\label{fig-energy2d}
    Localisation lengths as a function of energy for poly(dG)-poly(dC),
    telomeric, random-ATGC, and $\lambda$-DNA as described in the text.
    The spectrum is symmetric in energy. The data for telomeric DNA has
    been shaded for clarity. Lines are guides to the eye only.}
\end{figure}

Let us start by studying the localisation properties of DNA without any
onsite disorder either at $\varepsilon_{i,\tau}$ or at
$\varepsilon_{i,q}$.  For a poly(dG)-poly(dC) sequence, both fishbone
and ladder model produce two separate energy bands between the extremal
values computed at the end of section \ref{sec-effective}. Within these
energy bands, the electronic states are extended with infinite
localisation length $\xi$ as expected.  Outside the bands, transport is
exponentially damped due to an absence of states and the $\xi$ values
are very close the zero. In Fig.\ \ref{fig-inverse2d} the
resulting {\em inverse} localisation lengths are shown.
These are zero for the extended states in the two bands, but finite
outside, showing the quick decrease of the localisation lengths outside
the bands. 
In Fig.\ \ref{fig-energy2d}, we show the same data but now plot the
localisation length itself. 
We see that the energy gap observed previously \cite{CunCPD02} for the
poly(dG)-poly(dC) sequence in the fishbone model remains. The difference
with respect to the ladder model is a slight renormalisation of the gap
width.
The localisation lengths of poly(dG)-poly(dC) DNA tend to infinity,
meaning that the sequence is perfectly conducting. This is expected due
to its periodic electronic structure. 

Turning our attention to the other three DNA sequences, we find that
telomeric DNA also gives rise to perfect conductivity like
poly(dG)-poly(dC) DNA. But due to its structure of just 6 repeating base
pairs, there is a further split of each band into 3 separate sub-bands.
They may be calculated like in section \ref{sec-effective}.
We would like to point out that it may therefore be advantageous to use
the naturally occurring telomeric parts of DNA sequences as prime,
in-vivo candidates when looking for good conductivity in a DNA strand.

The structure of the energy dependence for the random-ATGC and the
$\lambda$-DNA is very different from the preceding two sequences, but it
is quite similar between just these two.  The biological content of the
DNA sequences is --- within the description by our quantum models ---
just a sequence of binary hopping elements between like and unlike base
pairs.  Thus the models are related to the physics of random hopping
models \cite{EilRS98a,BisCRS00} and in agreement with these, we see a
Dyson peak \cite{Dys53} in the centre of each sub-band.  Furthermore, we
see that the range of energies for which we observe non-zero
localisation lengths is increased into the gap and for large absolute
values of the energy. This is similar to the broadening of the single
energy band for the Anderson model of localisation \cite{RomS03}.
The localisation lengths, which roughly equal the average distance an
electron would be able to travel (conduct), are close to the distance of
$20$ bases within the band, with a maximum of $\sim 30$ bases at the
centre of each band. Note that this result is surprisingly good ---
given the level of abstraction used in the present models --- when
compared to the typical distances over which electron transfer processes
have been shown to be relevant
\cite{WanFSB00,BooLCD03,KelB99,MurAJG93,OneDB04,DelB03,TreHB01}.

\section{Results for disordered DNA}
\label{sec-results-disordered}

\subsection{DNA randomly bent or at finite temperatures}
\label{sec-uniform_energy}

As argued before, environmental influences on the transport properties
of DNA are likely to influence predominantly the electronic structure of
the backbone. Within our models, this can be captured by adding a
suitable randomness onto the backbone onsite potentials
$\varepsilon_{i}^{q}$. In this fashion, we can model for example the
influence of a finite-temperature \cite{BruGOR00} and thus a coupling to
phonons \cite{GutMC04}. We emphasise however, that in order for our
localisation results --- which rely on quantum mechanical interference
effects --- to remain valid, the phase breaking lengths should stay much
larger than the sequence lengths. Thus the permissible temperature range
is a few K only.
The bending of DNA is another possibility which can be modelled by a
local, perhaps regular, change in $\varepsilon_{i}^{q}$ along the strand.
Another important aspect is the change in $\varepsilon_{i}^{q}$ due to
the presence of a solution in which DNA is normally immersed.

\begin{figure}
  \centering
  \includegraphics[width=\figwidth]{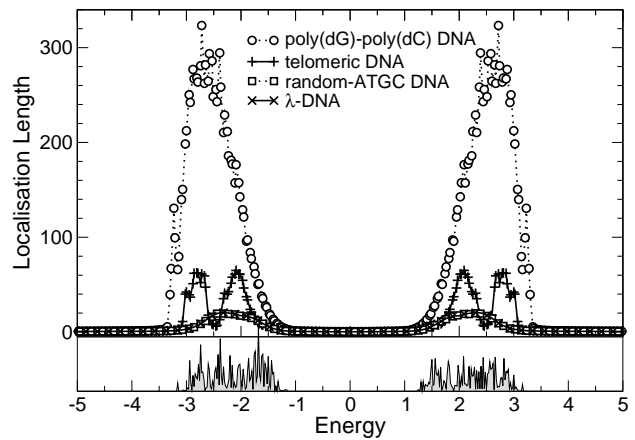}
  \caption{\label{fig-LM-uniW1-loc_energy}
    Top: Energy dependence of the localisation lengths, $\xi(E)$, for
    poly(dG)-poly(dC), telomeric, random-ATGC and $\lambda$-DNA in the
    presence of {\em uniform} backbone disorder with $W=1$. Only every
    2nd and 5th symbol is shown for random-ATGC and $\lambda$-DNA,
    respectively.
    Bottom: DOS for $\lambda$-DNA using the same parameters as in the
    top panel.}
\end{figure}

\begin{figure}
  \centering
  \includegraphics[width=\figwidth]{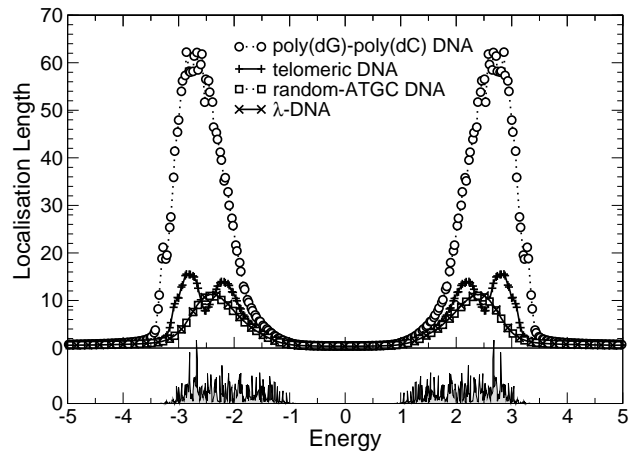}
  \caption{\label{fig-LM-uniW2-loc_energy}
    Top: $\xi(E)$ as in Fig.\ \ref{fig-LM-uniW1-loc_energy} but with
    $W=2$. Only every 2nd and 5th symbol is shown for random-ATGC and
    $\lambda$-DNA, respectively.
    Bottom: DOS for $\lambda$-DNA using the same parameters as in the
    top panel.}
\end{figure}

All these effects can be modelled in a first attempt by choosing an
appropriate distribution function $P(\varepsilon_{i}^{q}$). Let us first
choose uniform disorder with $\varepsilon_{i}^{q} \in [-W/2,W/2]$. 
In Fig.\ \ref{fig-LM-uniW1-loc_energy} we show the results for all 4 DNA
sequences as a function of energy for $W=1$. Comparing this to Fig.\ 
\ref{fig-energy2d}, we see that now all localisation lengths are finite;
poly(dG)-poly(dC) and telomeric DNA having localisation lengths of a few
hundreds and a few tens of bases, respectively. The localisation lengths
for random-ATGC and $\lambda$-DNA are only slightly reduced. In all
cases, the structure of 2 energy bands remains. Furthermore, $W=1$
already represents a sizable broadening of about $1/2$ the width of each
band. Thus although the localisation lengths are finite compared to the
results of section \ref{sec-results-clean}, they are still larger than
the lengths of the DNA strands used in the nano-electric experiments,
implying finite conductances. We remark that the Dyson peaks have
vanished as expected \cite{EilRS98a}. We plot the DOS for $\lambda$-DNA
in Fig.\ \ref{fig-LM-uniW1-loc_energy} which clearly indicates the $2$
bands.
Upon further increasing the disorder to $W=2$, as shown in Fig.\ 
\ref{fig-LM-uniW2-loc_energy}, the localisation lengths continue to
decrease. Note that we observe a slight broadening of the bands and
states begin to shift into the gap.
We also see that the behaviour of random-ATGC and $\lambda$-DNA is quite
similar and at these disorder strengths, even telomeric DNA follows the
same trends.
At $W=5$, the localisation lengths have been reduced to a few base-pair
separation distances and the differences between all $4$ sequences are
very small. The gap has been nearly completely filled as shown by the
DOS, albeit with states which have a very small localisation length.
This will become important later.

Thus, in summary, we have seen that adding uniform disorder onto the
backbone leads to a reduction of the localisation lengths and
consequently a reduction of the electron conductance. Strictly speaking,
all 4 strands are insulators. However, their localisation lengths can
remain quite large, larger than in many of the experiments. Thus even
the localised electron can contribute towards a finite conductivity for
these short sequences. In agreement with experiments, poly(dG)-poly(dC)
DNA is the most prominent candidate.
\begin{figure}
  \centering
  \includegraphics[width=\figwidth]{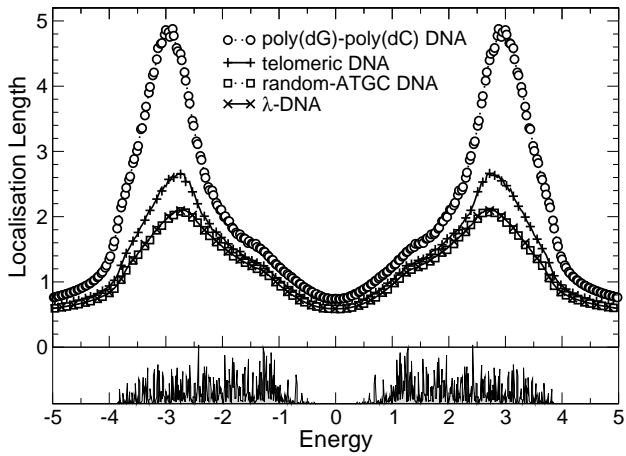}
  \caption{\label{fig-LM-uniW5-loc_energy}
    Top: $\xi(E)$ as in Fig.\ \ref{fig-LM-uniW1-loc_energy} but with
    $W=5$. Only every 2nd and 5th symbol is shown for random-ATGC and
    $\lambda$-DNA, respectively.
    Bottom: DOS for $\lambda$-DNA using the same parameters as in the
    top panel.}
\end{figure}

\subsection{DNA in an ionic solution}
\label{sec-binary_energy}

When in solution, the negatively charged oxygen on the backbone will
attract cations such as Na$^{+}$. This will give rise to a dramatic
change in local electronic properties at the oxygen-carrying backbone
site, but not necessarily influence the neighbouring sites. The effects
at each such site will be the same and thus in contrast to a uniform
disorder used in section \ref{sec-uniform_energy}, a binary distribution
such as $\varepsilon_{i,q}= \pm W/2$ is more appropriate.
For simplicity, we choose $50\%$ of all backbone sites to be occupied
$\varepsilon_{i,q}=-W/2$ while the other half remains empty with
$\varepsilon_{i,q}=+W/2$.  We note that a mixture of concentrations has
been studied in the context of the Anderson model in Ref.\ 
\cite{PlyRS03}.

\begin{figure}
  \centering
  \includegraphics[width=\figwidth]{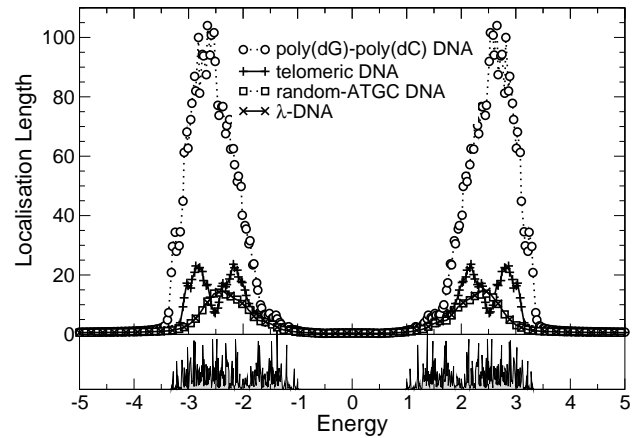}
  \caption{\label{fig-LM-binW1-loc_energy}
    Top: Energy dependence of the localisation lengths, $\xi(E)$, for
    poly(dG)-poly(dC), telomeric, random-ATGC and $\lambda$-DNA in the
    presence of {\em binary} backbone disorder with $W=1$. Only every
    2nd and 5th symbol is shown for random-ATGC and $\lambda$-DNA,
    respectively.
    Bottom: DOS for $\lambda$-DNA using the same parameters as in the
    top panel.}
\end{figure}

In Fig.\ \ref{fig-LM-binW1-loc_energy}, we show the results for moderate
binary disorder. In comparison with the uniformly disordered case of
Fig.\ \ref{fig-LM-uniW1-loc_energy}, we see that the localisation
lengths have decreased further. This is expected because binary disorder
is known to be very strong \cite{PlyRS03}. Also, the gap has already
started to fill.
%

\begin{figure}
  \centering
  \includegraphics[width=\figwidth]{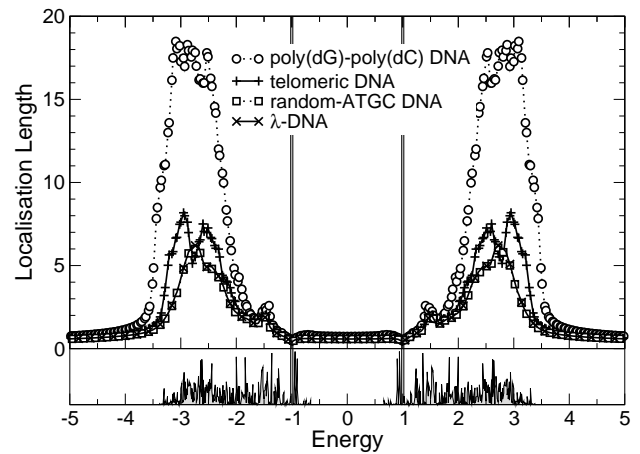}
  \caption{\label{fig-LM-binW2-loc_energy}
    Top: $\xi(E)$ as in Fig.\ \ref{fig-LM-binW1-loc_energy} but with
    $W=2$. Only every 2nd and 5th symbol is shown for random-ATGC and
    $\lambda$-DNA, respectively.
    Bottom: DOS for $\lambda$-DNA using the same parameters as in the
    top panel.}
\end{figure}

\begin{figure}
  \centering
  \includegraphics[width=\figwidth]{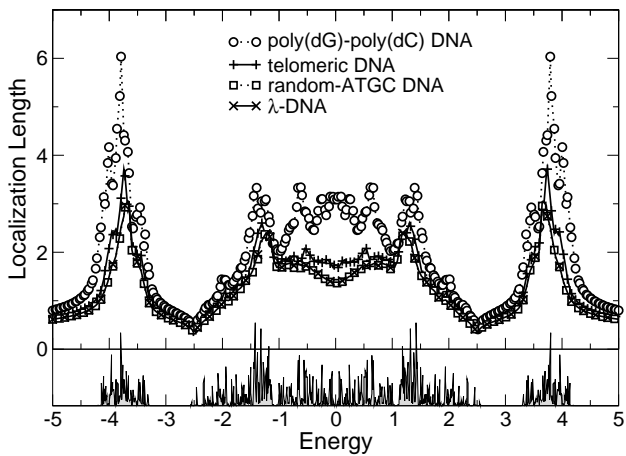}
  \caption{\label{fig-LM-binW5-loc_energy}
    Top: $\xi(E)$ as in Fig.\ \ref{fig-LM-binW1-loc_energy} but with
    $W=5$. Only every 2nd and 5th symbol is shown for random-ATGC and
    $\lambda$-DNA, respectively.
    Bottom: DOS for $\lambda$-DNA using the same parameters as in the
    top panel.}
\end{figure}

Increasing the disorder leads again to a decrease of $\xi$ in the energy
regions corresponding to the bands. Directly at $E=\pm W/2$, we observe
$2$ strong peaks in the DOS which is accompanied by reduced localization
lengths. This peak corresponds to the infinite potential barrier or well
at $E=-W/2$ or $+W/2$, respectively, as indicated by Eq.\ 
(\ref{eq-ladder-effective}).  In Fig.\ \ref{fig-LM-binW1-loc_energy},
these peaks were not yet visible. We also see in Fig.\ 
\ref{fig-LM-binW2-loc_energy} that the localisation lengths for states
in the band centre start to increase to values $\gtrsim 1$.
This trend continues for larger $W$ as shown in Fig.\ 
\ref{fig-LM-binW5-loc_energy}. We see a crossover into a regime where
the two original, weak-disorder bands have nearly vanished and states in the
centre at $E=0$ are starting to show an increasing localisation length
{\em upon increasing the binary disorder}.
A further increase in $W$ eventually leads to the complete destruction
of the original bands and the formation of a single band symmetric
around $E=0$ at about $W\sim 2.5$.

\subsection{Delocalisation due to disorder}
\label{sec-delocalization}

\begin{figure}
  \centering
  \includegraphics[width=\figwidth]{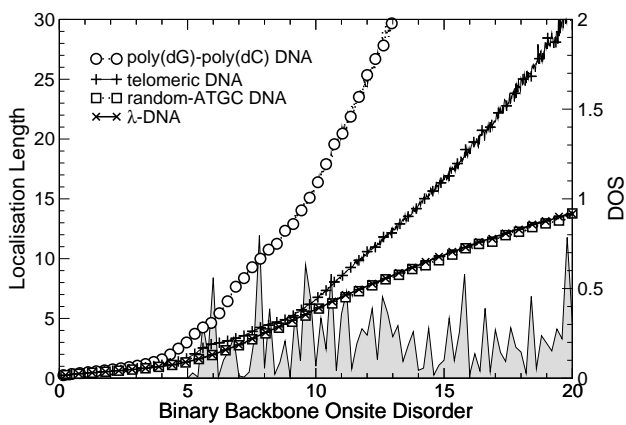}
  \caption{\label{fig-LM-binE0-loc_disorder}
    Disorder dependence of $\xi$ for poly(dG)-poly(dC), telomeric,
    random-ATGC and $\lambda$-DNA at $E=0$. Only every 10th symbol is
    shown for all sequences. The shaded curve is the corresponding unnormalized DOS
    for $\lambda$-DNA.}
\end{figure}

\begin{figure}
  \centering
  \includegraphics[width=\figwidth]{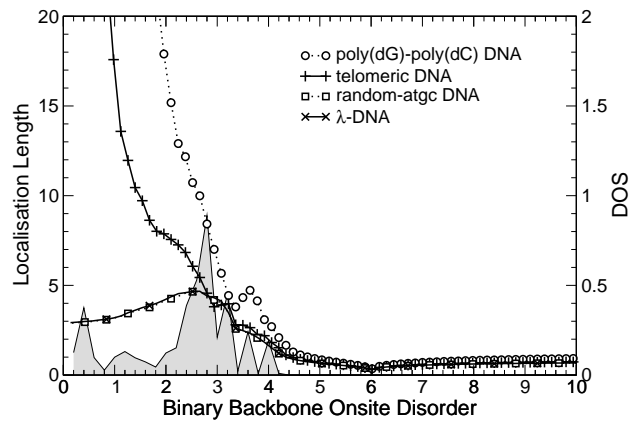}
  \caption{\label{fig-LM-binE3-loc_disorder}
    $\xi(W)$ as in Fig.\ \ref{fig-LM-binE0-loc_disorder} but with $E=3$.
    Only every 10th symbol is shown for all DNA sequences. The shaded
    curve is the corresponding unnormalized DOS for $\lambda$-DNA.}
\end{figure}

\begin{figure}
  \centering
  \includegraphics[width=\figwidth]{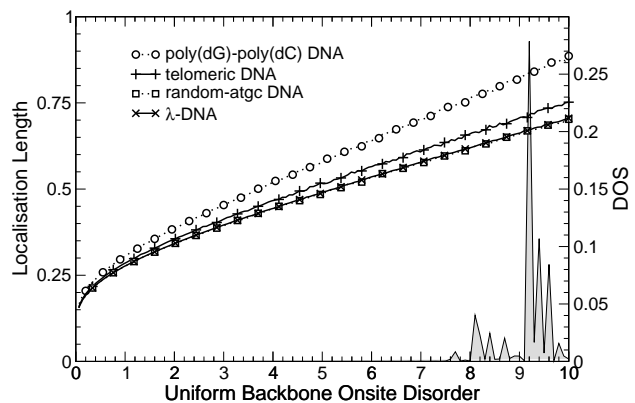}
  \caption{\label{fig-FM-uniE0-loc_disorder}
    $\xi(W)$ as in Fig.\ \ref{fig-LM-binE0-loc_disorder} but with
    uniform disorder at $E=0$ and for the {\em fishbone model}.  Only
    every 10th symbol is shown for all DNA sequences. The shaded curve
    is the corresponding unnormalized DOS for $\lambda$-DNA.}
\end{figure}

The results of the previous section suggest that increasing the disorder
in different regions of the energy will lead to different transport
behaviour. Of particular interest is the region at $E=0$. In Fig.\ 
\ref{fig-LM-binE0-loc_disorder} the variation of $\xi$ as a function of
binary disorder strength for all different sequences is shown.  While
$\xi < 1$ for small disorder, we see that upon increasing the disorder,
states begin to appear and their localisation lengths increase for all
DNA sequences. Thus we indeed observe a counter-intuitive {\em
  delocalisation} by disorder at $E=0$. As before, poly(dG)-poly(dC) and
telomeric disorder show the largest localisation lengths, whereas
random-ATGC and $\lambda$-DNA give rise to a smaller and nearly
identical effect.
In Fig.\ \ref{fig-LM-binE3-loc_disorder} we show that this effect does
not exist at $E=3$, i.e.\  for energies corresponding to the formerly
largest localisation lengths. Rather, at $E=3$, the localisation lengths
for all DNA sequences quickly drop to $\xi \sim 1$.
The delocalisation effect is also observed for uniform disorder, but is
much smaller. As shown in Fig.\ \ref{fig-FM-uniE0-loc_disorder}, the
enhancement is up to about $\xi=1$ for the fishbone model
(\ref{eq-fishbone}). Results for the ladder model (\ref{eq-ladder}) are
about $1.7$ times larger.

This surprising delocalisation-by-disorder behaviour can be understood
by considering the effects of disorder at the backbone for the effective
Hamiltonians (\ref{eq-fishbone-effective}) and
(\ref{eq-ladder-effective}). At $E=0$, the onsite potential correction
term ${\left(t_{i}^{q}\right)^{2}}/{(\varepsilon_{i}^{q} - E)}$ will
{\em decrease} upon increasing the $\varepsilon_{i}^{q}$ values. For
binary disorders $\varepsilon_{i}^{q} = \pm W/2$, this holds for
$|\varepsilon_{i}^{q}| > |E|$ as shown in Fig.\ 
\ref{fig-LM-binE3-loc_disorder}. However, for large $|E|$, the
localisation lengths decrease quickly due to the much smaller density of
states. Thus the net effect is an eventual decrease (or an only very
small increase) of $\xi$ for large $E$. Note the dip at
$|\varepsilon_{i}^{q}|=E=3$ in the figure, which corresponds to the
effective $\varepsilon_{i}= \infty$, i.e.\ an infinitely strong trap
yielding extremely strong localisation.
For uniform disorder $\varepsilon_{i}^{q} \in [-W/2,W/2]$ --- and
generally any disorder with compact support around $E=0$ --- the above
inequality is never full-filled and even for $E=0$ we will find small
$\varepsilon_{i}^{q} \sim 0$ such that we have strong trapping and
localisation.

\section{Investigating the local properties of the sequences}
\label{sec-local}

\subsection{Variation of $\xi$ along the DNA strand}

In the preceding sections, we had computed estimates of the localisation
length $\xi$ for complete DNA strands, i.e.\ the $\xi$ values are {\em
  averages}. However, the biological function of DNA clearly depends on
the local structure of the sequence in a paramount way. After all, only
certain parts of DNA code for proteins, while others do not. In
addition, the exact sequence of the bases specifies the protein that is
to be assembled.
Thus, in order to gain access to the local properties, we have performed
computations of $\xi$ on subsequences of complete DNA strands. We start
by artificially restricting ourselves to finite windows of length $K=
10, 30, 50, 100, 200, 500, 1000$ and compute the localisation lengths
$\xi_{K}(r)$ where $r=1, 2, \ldots, L-K$ denotes the starting position
of the window of length $K$.

In order to see how the exact sequence determines our results, we have
also randomly permuted (scrambled) the $\lambda$-DNA sequence so that
the content of A, T, G, and C bases is the same, but their order is
randomised. Differences in the localisation properties should then
indicate the importance of the exact order. 
From the biological information available on bacteriophage
$\lambda$-DNA, we compute the localisation length for the coding regions
\cite{DanSSS83} and then for window lengths $K$ that correspond exactly
to the length of each coding region. Again, if the electronic properties
--- as measured by the localisation length --- are linked to biological
content, we would expect to see characteristic differences.

\begin{figure}
  \centering
  \includegraphics[width=\figwidth]{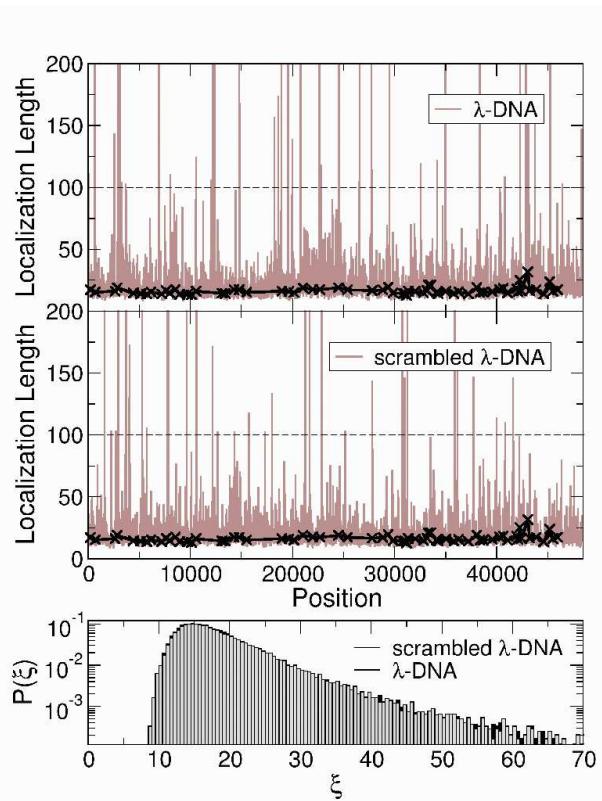}
  \caption{\label{fig-LM-cleanE3-loc_window-100}
    Top: Variation of the localisation lengths for a sliding window of
    length $K=100$ as a function of window starting position for
    $\lambda$-DNA at $E=3$. The black crosses ($\times$) denote results
    for windows corresponding to the coding sequences of $\lambda$-DNA
    only. The dashed horizontal line denotes $K$.
    Middle: Same as in the top panel but with randomly scrambled
    $\lambda$-DNA.
    Bottom: Normalised distribution functions $P(\xi)$ for the
    localisation lengths $\xi$ of $\lambda$- (black) and
    scrambled-$\lambda$-DNA (grey). }
\end{figure}

\begin{figure}
  \centering
  \includegraphics[width=\figwidth]{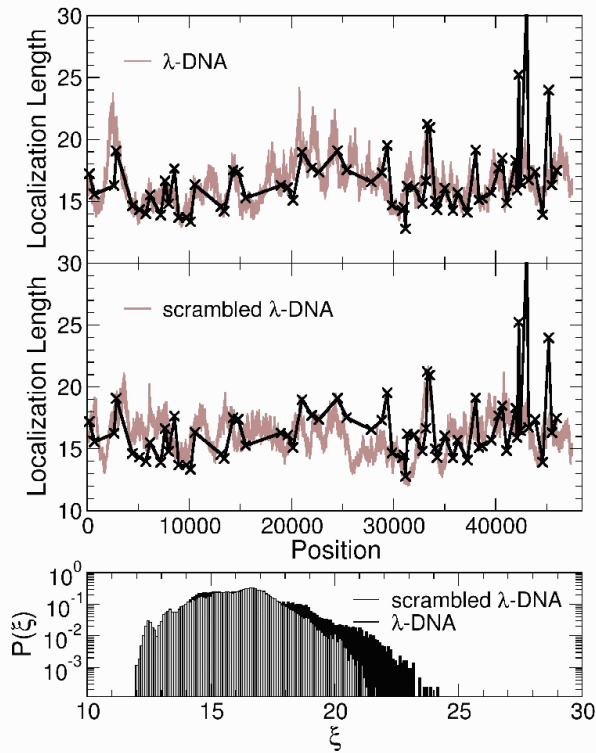}
  \caption{\label{fig-LM-cleanE3-loc_window-1000}
    Variation of the localisation lengths for a sliding window of length
    $K=1000$ at $E=3$ as in Fig.\ 
    \protect\ref{fig-LM-cleanE3-loc_window-100}.
    Middle: Same as in the top panel but with randomly scrambled
    $\lambda$-DNA.
    Bottom: Normalised distribution functions $P(\xi)$ for the
    localisation lengths $\xi$ of $\lambda$- (black) and
    scrambled-$\lambda$-DNA (grey). }
\end{figure}

In Figs.\ \ref{fig-LM-cleanE3-loc_window-100} and
\ref{fig-LM-cleanE3-loc_window-1000}, we show results for $K=100$ and
$1000$, respectively.
From Fig.\ \ref{fig-LM-cleanE3-loc_window-100}, we see from $P(\xi)$
that the localisation lengths for $\lambda$-DNA are mostly distributed
around $15$--$20$, but $P(\xi)$ has a rather long tail for large $\xi$.
However, there are some windows where the localisation lengths exceed
even the size of the window $K=100$. Thus at specific positions in the
DNA sequence, the system appears essentially extended with $\xi > K$. On
the other hand, the distribution $P(\xi)$ is identical when instead of
$\lambda$-DNA, we consider scrambled DNA.  Therefore the presence of
such regions is not unique to $\lambda$-DNA.  The results from windows
positioned at the coding part of $\lambda$-DNA appear statistically
similar to the complete sequence, i.e.\ including also the non-coding
regions. This suggests that with respect to the localisation properties
there is no obvious difference between $\lambda$-DNA and scrambled
$\lambda$-DNA as well as coding and non-coding regions. We emphasise
that similar results have been obtained for a DNA sequence constructed
from the SARS corona-viral data.

In Fig.\ \ref{fig-LM-cleanE3-loc_window-100}, we repeat these
calculations but with $K=1000$. Clearly, $P(\xi)$ is peaked again around
$15$--$20$ and this time has no tail. In all cases, $K>\xi$. Again, the
results for scrambled DNA are different in each window, and now even
$P(\xi)$ is somewhat shifted with respect to $\lambda$-DNA.

Thus in conclusion, we do not see significant differences between
$\lambda$-DNA and its scrambled counter part. Moreover, there appears to
be no large difference between the localisation lengths measured in the
coding and the non-coding sequences of bacteriophage $\lambda$-DNA. This
indicates that the average $\xi$ values computed in the previous
sections is sufficient when considering the electronic localisation
properties of the $4$ complete DNA sequences.

\subsection{Computing correlation functions}

As shown in the last section, the spatial variation of $\xi$ for a fixed
window size is characteristic of the order of bases in the DNA sequence.
Thus we can now study how this biological information is retained at the
level of localisation lengths. In order to do so, we define the
correlation function
\begin{equation}
{\rm Cor}(k)
=
\frac{\sum_{i=1}^{n-k}\left[\xi(r_i)-\langle{\xi}\rangle\right]
\left[\xi(r_{i+k})-\langle{\xi}\rangle\right]}%
{\sum_{i=1}^{n}\left[\xi(r_i)-\langle{\xi}\rangle\right]^2}
\label{eq-cor}
\end{equation}
where $\langle{\xi}\rangle={\sum_{i=1}^{n}\xi(r_i)}/{n}$ is $\xi$
averaged over all $n=L-(K-1)$ windows for each of which the individual
localisation lengths are $\xi(r_i)$.

\begin{figure}
  \centering
  \includegraphics[width=\figwidth]{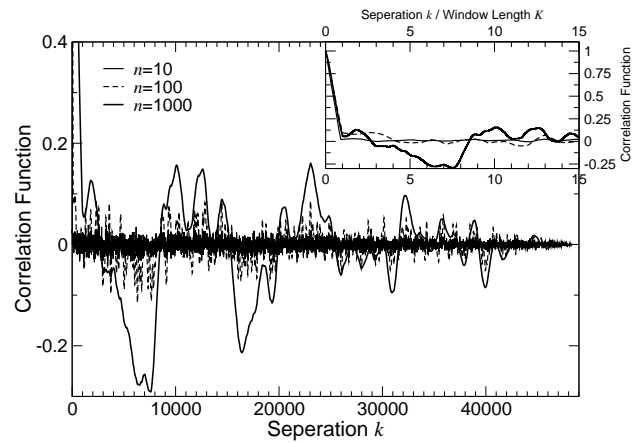}
  \caption{\label{fig-LM-Win-E0-cor_pos}
    ${\rm Cor}(k)$ as defined in Eq.\ (\ref{eq-cor}) for $\lambda$-DNA and
    $K=10$, $100$, and $1000$ at $E=0$. The inset shows the same date
    but plotted as a function of normalized seperation $k/K$.}
\end{figure}

In Fig.\ \ref{fig-LM-Win-E0-cor_pos} we show the results obtained for
$\lambda$-DNA with windows of length $10$, $100$ and $1000$. We first
note that ${\rm Cor}(k)$ drops rapidly until the distance $k$ exceeds
the window width $K$ (see the inset of Fig.\ 
\ref{fig-LM-Win-E0-cor_pos}). For $k>K$, ${\rm Cor}(k)$ fluctuates
typically between $\pm 0.2$ and there is a larger anti-correlation for
base-pair separations of about $k=8000$. We note that such large scale
features are not present when considering scrambled $\lambda$-DNA
instead.

\section{Discussion}

The fishbone and ladder models studied in the present paper give
qualitatively similar results, i.e.\ a gap in the DOS on the order of
the hopping energies to the backbone, extended states for periodic DNA
sequences and localised states for any non-zero disorder strength. Thus
at $T=0$, our results suggest that DNA is an insulator unless perfectly
ordered.  Quantitatively, the localisation lengths $\xi$ computed for
the ladder model are larger than for the fishbone model.  Since we are
interested in these non-universal lengths, the ladder model is clearly
the more appropriate model.


The localisation lengths measure the spatial extent of a conducting
electron. Our results suggest --- in agreement with all previous
considerations --- that poly(dG)-poly(dC) DNA allows the largest values
of $\xi$. Even after adding a substantial amount of disorder,
poly(dG)-poly(dC) DNA can still support localization lengths of a few
hundred base-pair seperation lengths. With nanoscopic experiments
currently probing at the most a few dozen bases, this suggests that
poly(dG)-poly(dC) DNA will appear to be conducting in these experiments.

Furthermore, telomeric DNA is a very encouraging and interesting
naturally occuring sequence because it gives very large localisation
lengths in the weakly disordered regime.  Nevertheless, we find that all
investigated, non-periodic DNA sequences such as, e.g.\ random-ATGC and
$\lambda$-DNA, give localised behaviour even in the clean state. This
indicates that they are insulating at $T=0$.


When the effects of the environment, modelled by their potential changes
on the backbone, are included, we find that the localisation lengths in
the two bands decrease quickly upon increasing the disorder.
Nevertheless, depending on the value of the Fermi energy, the resulting
$\xi$ values can still be 10-20 base-pairs long.  While this may not
give metallic behavior, it can still result in a finite current for
small sequences. We also note that these distances are quite close to
those obtained from electron-transfer studies.


The backbone disorder also leads to states moving into the gap.
Therefore the environment prepared in the experiments determines the gap
which is being measured. Furthermore, the localisation properties of the
states in the former gap are drastically different from those in the 2
bands.  Increasing the disorder leads to an increase in the localization
lengths and thus potentially larger currents. This is most pronounced
for binary disorder, taken to model the adhesion of cations in solution.
Thus within the $2$ models studied, we find that their transport
properties are in a very crucial way determined by the environment.
Differences in experimental set-up such as measurements in 2D surfaces
or between elevated contacts are likely to lead to quite different
results.

%

As far as the correlations within biological $\lambda$-DNA are
concerned, we see only a negligible difference between the localisation
properties of the coding and non-coding parts. However, this is clearly
dependent on the chosen energy and the particular window lengths used.
Investigations on other DNA sequences are in progress.

\acknowledgments
It is a pleasure to thank H.\ Burgert, D.\ Hodgson, M.\ Pfeiffer and D.\ 
Porath for stimulating discussions.


\begin{thebibliography}{60}
\bibitem{perpendiculartolongaxis} We note that Ref.\ \cite{WalZ04} assumes
transport is via the sugar-phosphate backbone.
\bibitem{fishbone} The results for the fishbone and ladder models are
qualitatively the same. Quantitatively, the ladder model results have a
nearly twice larger localisation length. This factor approaches $2$, if
$t_{1,2}\rightarrow 0$. Therefore we will focus our discussion on the
2-channel ladder model.
\end{thebibliography}

\end{document}